\documentclass[fleqn,usenatbib]{mnras}

\usepackage{newtxtext,newtxmath}
\usepackage{lineno}

\usepackage[T1]{fontenc}

\DeclareRobustCommand{\VAN}[3]{#2}
\let\VANthebibliography\thebibliography
\def\thebibliography{\DeclareRobustCommand{\VAN}[3]{##3}\VANthebibliography}

\newcommand{\burke}{\citetalias{2021Sci...373..789B}}
\newcommand{\ren}{\citetalias{2024ApJ...975..160R}}
\defcitealias{2021Sci...373..789B}{B21}
\defcitealias{2024ApJ...975..160R}{R24}

\usepackage{graphicx}	
\usepackage{amsmath}	

\title[BADDAT]{Baseline-Aware Dependence fitting for DAmping Timescales (BADDAT): A Nearly Unbiased Approach to Constraining Optical Variability Dependence on Physical Properties of Active Galactic Nuclei}

\author[Xia et al.]{
Ruisong Xia$^{1,2}$\thanks{E-mail: xiars@mail.ustc.edu.cn},
Zhen-Yi Cai$^{1,2}$\thanks{E-mail: zcai@ustc.edu.cn},
Yongquan Xue$^{1,2}$\thanks{E-mail: xuey@ustc.edu.cn},
Xian-Liang Lu$^{3}$,
Guowei Ren$^{1,2}$,
Shuying Zhou$^{4}$,
\newauthor
Mouyuan Sun$^{4}$,
Shifu Zhu$^{1,2}$,
Zhen-Bo Su$^{1,2}$,
and Hao Liu$^{1,2}$,
\\
$^{1}$Department of Astronomy, University of Science and Technology of China, Hefei 230026, China\\
$^{2}$School of Astronomy and Space Science, University of Science and Technology of China, Hefei 230026, China\\
$^{3}$School of Physics, Sun Yat-sen University, Guangzhou 510275, China\\
$^{4}$Department of Astronomy, Xiamen University, Xiamen, Fujian 361005, China\\
}

\date{Accepted XXX. Received YYY; in original form ZZZ}

\pubyear{\the\year{2025}}

\begin{document}
\label{firstpage}
\pagerange{\pageref{firstpage}--\pageref{lastpage}}
\maketitle

\begin{abstract}
Active galactic nuclei (AGNs) exhibit stochastic optical variability, commonly characterized by a damped random walk.
The damping timescale is of particular interest because it is related to fundamental properties of the central black hole, such as its mass and accretion rate.
However, the systematic underestimation of damping timescales caused by limited observational baselines makes it difficult to exhaustively utilize all available data.
Many previous efforts have relied on strict selection criteria to avoid biased measurements, and such criteria inevitably constrain the range of AGN physical parameter space and therefore hinder robust inference of the underlying dependencies of damping timescale on AGN properties.
In contrast, we introduce a novel forward modeling approach, Baseline-Aware Dependence fitting for DAmping Timescales (BADDAT), which explicitly accounts for these biases and leverages the information contained in underestimated timescale measurements. 
Rather than attempting to correct individual timescale measurements, BADDAT robustly constrains the population-level dependence of damping timescale on AGN physical properties.
We demonstrate its effectiveness using mock light curves and show that it successfully reconciles previous inconsistent results based on two independent AGN samples.
Our BADDAT method will have broad applications in AGN variability studies during the era of time-domain astronomy.

\end{abstract}

\begin{keywords}
{accretion, accretion discs -- galaxies: active -- galaxies: nuclei -- galaxies: supermassive black holes -- methods: data analysis}
\end{keywords}



\section{introduction}
Active galactic nuclei (AGNs) exhibit significant variability driven by extreme physical processes occurring in the accretion disks surrounding their central supermassive black holes. 
While some AGNs exhibit quasiperiodic features, the majority of their variability is stochastic.
Damped random walk (DRW), the simplest form of Gaussian processes, has been widely and effectively used to model the stochastic optical variability of AGNs with two fundamental variability parameters, i.e., a damping timescale $\tau$ and a variability amplitude $\sigma$ \citep[e.g.,][]{2009ApJ...698..895K,2010ApJ...708..927K,2010ApJ...721.1014M, 2011ApJ...727L..24D, 2013ApJ...765..106Z,2021ApJ...907...96S}.
The correlation between the measured timescale and physical parameters {of the central black holes} is significant across multiple samples \citep[e.g.,][]{2009ApJ...698..895K, 2010ApJ...721.1014M, 2021Sci...373..789B, 2022MNRAS.514..164S, 2024ApJ...975..160R,2024ApJ...967L..18Z}.

It is common practice to impose a limit on the estimated timescale $\tau_{\rm DRW}$ relative to the cadence and baseline, although this approach can be quite restrictive.
For example, \citet{2017A&A...597A.128K} stated that the typical measured timescale should be less than $\sim20\%$ of the baseline (i.e., the time span of the light curve).
\citet{2021Sci...373..789B} required the measured timescale to be less than $10\%$ based on their simulations.
\citet{2024ApJ...961....5H} exhaustively explored the fitting methods and confirmed that the {timescale} can be well retrieved when {it} is less than $10\%$ of the baseline.
{\citet{2024ApJ...966....8Z} further emphasized that it is the intrinsic timescale rather than the measured timescale that must be less than $10\%$ of the baseline for the estimates to remain unbiased.}

\citet{2024ApJ...975..160R} applied the criterion that the theoretical damping timescale $\tau_{\rm th}$, {representing the intrinsic timescale} as predicted by physical models, must be less than $10\%$ of the baseline, rather than selecting based on the measured timescale.
If the intrinsic timescale can be accurately predicted by theory, this approach can effectively exclude measurements affected by insufficient light-curve lengths.  
However, since this criterion is stringent and excludes the majority of light curves in the sample, it substantially restricts the parameter space over which the {linear regression for the properties can be evaluated}.
Consequently, the results may still be biased during regression due to the intrinsic scatter, even if the timescale estimates for the remaining light curves are perfectly accurate.

\begin{figure}
    \centering
    \includegraphics[scale=0.5]{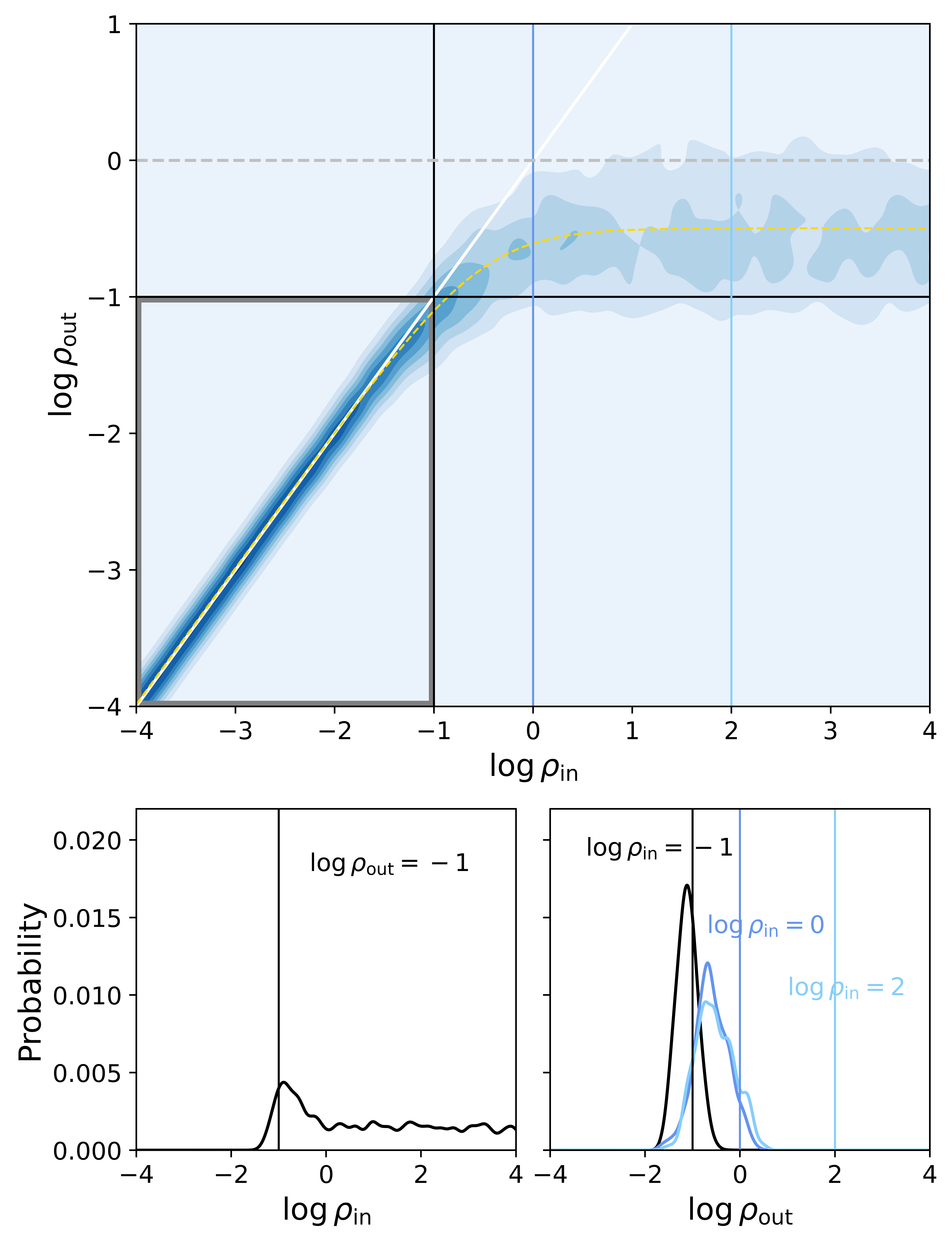}
    \caption{Probability distribution of the input and output $\rho = \tau/\text{baseline}$ on a logarithmic scale.
    {The mock light curves, each spanning 10000 days and containing 10000 data points, are generated using the DRW model with input timescales in the range of $-3 < \log (\tau_{\mathrm{in}}/\mathrm{days}) < 9$ and input amplitudes in the range $-2 < \log \sigma_{\rm in} < 2$. Both parameters are evenly sampled in logarithmic space, with steps of 0.01 and 0.04, respectively. We select light curves satisfying $\log (\sigma_{\rm out}/0.3) < 0.1$, ensuring that the output amplitude remains close to 0.3, resulting in a total of 4741 mock light curves for this demonstration.
    }
    The upper panel shows the joint probability distribution map.
    The {horizontal} dashed line marks $\log \rho_{\rm out} = 0$, which appears to be a soft boundary for $\rho_{\rm out}$.
    The lower panels show the conditional probability distributions {of} $P(\log \rho_{\rm in} | \log \rho_{\rm out} = -1)$ and $P(\log \rho_{\rm out} | \log \rho_{\rm in} = -1)$ {in black}.
    The black solid lines mark $\log \rho_{\rm in} = -1$ and $\log \rho_{\rm out} = -1$ in all panels, which correspond to the selection criteria from \citet{2021Sci...373..789B} and \citet{2024ApJ...975..160R}, respectively. 
    The gray rectangle marks the region considered by both, while the outer region represents the parameter space that BADDAT can encompass, significantly expanding the {parameter} space where most of the samples are located.
    The lighter blue lines and curves illustrate the cases with $\log \rho_\mathrm{in} = 0$ and $\log \rho_\mathrm{in} = 2$.
    {The yellow dashed line represents the modeled result of $\xi(\log \rho)$.}
    }
    \label{fig:mle}
\end{figure}

As widely simulated and discussed \citep[e.g.,][]{2009ApJ...698..895K, 2017A&A...597A.128K, 2021Sci...373..789B, 2024ApJ...961....5H}, the mechanism of timescale underestimation has been extensively investigated, with underestimated timescales typically falling within the range of 0.1 to 1 times the baseline value {when observational baselines are limited.
After correcting for measurement biases introduced by observational noise with the recently developed SIMEX algorithm \citep{2025MNRAS.540..521E}, the pattern of underestimation is essentially unchanged.}
There is potential to extract valuable information by incorporating these underestimated data {arising from limited observational baselines} into the analysis.

Here, we do not aim to optimize selection strategies to obtain unbiased measurements of the timescale for each individual light curve, but rather focus on more robustly constraining the dependence of timescales on physical properties. 
We demonstrate that the dependence can be better constrained if underestimated data are included in the analysis and the related biases are properly accounted for via a forward modeling technique.
We refer to this method as Baseline-Aware Dependence fitting for DAmping Timescales (BADDAT).
BADDAT deals with the ``bad data'' ({see} Section~\ref{sec:meth}) and yields nearly unbiased results ({see} Section~\ref{sec:test}), providing a robust framework for observations and future studies ({see} Section~\ref{sec:imp}).

\section{Methodology}\label{sec:meth}

\subsection{How Damping Timescale {Is} Underestimated}\label{sec:map}

Maximum likelihood estimation (MLE) is one of the simplest and most efficient methods for obtaining the DRW parameters from time series.
Here, we revisit the issue of damping timescale underestimation using MLE.

We generate a set of mock light curves based on the DRW model, with input timescales in the range {of} $-3 < \log (\tau_{\mathrm{in}}/\mathrm{days}) < 9$ and input amplitudes in the range {of} $-2 < \log \sigma_{\rm in} < 2$, empirically covering the full extent of reasonable $\tau$ and $\sigma$ values observed in AGNs \citep[e.g.,][]{2010ApJ...721.1014M, 2021Sci...373..789B, 2022MNRAS.514..164S, 2024ApJ...975..160R}. 
Both parameters are evenly sampled in logarithmic space, with steps of 0.01 and 0.04, respectively, {resulting in 121301 mock light curves}. 
The baseline is set to 10000 days with 10000 data points in each light curve.
In this case, the cadence, defined as the mean interval between data points, is 1 day, and the {10000} sampling times are randomly drawn from a uniform distribution over the baseline.
The uncertainties of the mock light curves are assigned a value of {0.1 magnitude}.

We fit the mock light curves with \textsc{celerite} \citep{celerite} in an MLE manner based on the \textsc{taufit} code (https://github.com/burke86/taufit; \citealt{2021Sci...373..789B}) and obtain the output damping timescale $\tau_{\rm out}$ and variability amplitude $\sigma_{\rm out}$ for each $\tau_{\rm in}$ and $\sigma_{\rm in}$.
Following the criteria adopted by \cite{2024ApJ...975..160R}, we use the Akaike Information Criterion (AIC; \citealt{1974ITAC...19..716A}) to evaluate the fitting results, where $\text{AIC} = -\ln(\mathcal{L}) + 2N$, with $\mathcal{L}$ being the likelihood {of the} MLE and $N$ the number of parameters.
The AIC of each fit {($\text{AIC}_{\rm best}$)} is compared with $\text{AIC}_{\rm low}$ and $\text{AIC}_{\rm hi}$ of the alternative {fits of \hbox{\(\tau_{\rm out}= \text{cadence}/100\)} and \hbox{\(\tau_{\rm out}= 100\times \text{baseline}\)}} as extremely small and large values, respectively.
Considering a less strict criterion, we experimentally exclude fitting failures if $\Delta \text{AIC}_{\rm low} = \text{AIC}_{\rm low} - \text{AIC}_{\rm best}$ and $\Delta \text{AIC}_{\rm hi} = \text{AIC}_{\rm hi} - \text{AIC}_{\rm best}$ are either less than 1, {indicating that the best-fit model is not substantially better than the unreasonable alternatives.}
We select light curves satisfying $\log (\sigma_{\rm out}/0.3) < 0.1$, {resulting in a total of 4741 light curves}, {which} ensures that the output amplitude is close to 0.3, corresponding to a typical signal-to-noise ratio (SNR) observed in real AGN light curves under an assumed noise level of 0.1 magnitude, {where SNR is defined as the ratio of the amplitude of the DRW to the noise \citep{2021Sci...373..789B}.}
This selection allows us to derive the joint probability distribution $P(\log \tau_{\rm in}, \log \tau_{\rm out})$ {within a certain range of $\sigma_{\rm out}$}.
{This distribution encapsulates the full statistical dependence between $\tau_{\rm in}$ and $\tau_{\rm out}$, providing a comprehensive basic knowledge for subsequent inference and interpretation.}

The shape of the joint probability distribution does not change significantly if we alter the selected $\sigma_{\rm out}$ within a reasonable range.
However, the probability distribution depends on the sampling pattern, which differs in every light curve.
It is impossible to simulate all the situations for each light curve due to computational limitations.
As a compromise, we repeat the simulations with a different number of data points, i.e., uniformly choosing 10 values from 10 to 10000 days on a logarithmic scale. 
{All the resulting distributions converge to a clear shape} (similar to the upper panel of Figure~\ref{fig:mle}, {showing the case with baseline of 10000 days}) as the number of data points {becomes} large.
The difference between two adjacent maps is not significant.
Therefore, when analyzing a specific light curve, we approximate its probability distribution using the precomputed map corresponding to the closest available number of data points.

As an example, the upper panel of Figure~\ref{fig:mle} shows the probability density map for {light curves with 10000 data points.}
The timescale is divided by the assumed baseline, $\rho = \tau / \text{baseline}$, for extended use.
The range where either $\tau_{\rm in} < \text{cadence}$ or $\tau_{\rm out} < \text{cadence}$ is not shown.
Similar to previous works, light curves of insufficient {lengths (i.e., $\log \rho_{\rm in}>-1$)} lead to bias \citep[e.g.,][]{2009ApJ...698..895K, 2017A&A...597A.128K,  2021Sci...373..789B, 2024ApJ...961....5H}.
The conditional probability distributions {of} $P(\log \rho_{\rm in}|\log \rho_{\rm out}=-1)$ and $P(\log \rho_{\rm out}|\log \rho_{\rm in}=-1)$ are shown in the lower panels of Figure~\ref{fig:mle}, respectively.
{It is clear that} although when $\log \rho_{\rm in}=-1$, i.e., the input timescale equals 0.1 times of baseline, the output timescale is distributed like a Gaussian, while $P(\log \rho_{\rm in}|\log \rho_{\rm out}=-1)$ shows a long tail to larger timescales, indicating {that} the estimation is most likely underestimated.

{Previous works \citep[e.g.,][]{2021Sci...373..789B, 2024ApJ...975..160R} exclude data deemed to be underestimated, but such stringent selection criteria may introduce additional bias (see in Section~\ref{sec:test}) by systematically removing a specific region from an intrinsically scattered distribution.
With an understanding of how the timescale $\tau_{\rm out}$ tends to be underestimated ({as shown in} Figure~\ref{fig:mle}), we can incorporate the underestimated data in a statistically informed manner, enabling more reliable and informative inference of the timescale dependencies, as detailed in the following subsection.}

\subsection{BADDAT: an Approach to Quantifying Parameter Dependence}\label{sec:app}

Motivated by the desire to fully utilize the possibly underestimated timescales, we propose designing an approach to quantifying their dependence on different physical parameters with looser criteria and considering more data. 
{Essentially, given that the probability density function of the underestimation is known (see Section~\ref{sec:map}), BADDAT incorporates underestimated data through modified Bayesian linear regression of timescales.}
In this approach, we model the logarithm of the unbiased timescale \(\tau_{{\rm in},i}\) as a linear function of the independent variables \(X_{i,j}\), written in terms of \(\rho_{{\rm in},i}\):
\begin{equation}
    \log (\rho_{{\rm in},i}) =\sum_j k_j X_{i,j} + b + \log(1+z_i) - \log(\text{baseline}_i), 
\end{equation}
where the third term in the right part of the equation accounts for the effect of redshift \(z_i\) on the timescale.

The posterior probability function of the linear coefficient {$k_j$} can be written as:
\begin{equation}\label{eqo}
    P(\{k_j\},b|\log \rho_{{\rm out},i}) = \frac{P(\log \rho_{{\rm out},i}|\{k_j\},b)P(\{k_j\}, b)}{P(\log \rho_{{\rm out},i})},
\end{equation}
where {\(\{k_j\}\) represents the entire set of \(k_j\)}, the likelihood function \( P(\log \rho_{{\rm out},i} | \{k_j\},b) \) quantifies how well the model parameters explain the observed data, the prior \( P(\{k_j\},b) \) encodes our beliefs about the model parameters, and \( P(\log \rho_{{\rm out},i}) \) serves as a normalizing constant and can be omitted when searching for the maximum posterior.

The likelihood for this model can be written as:
\begin{equation}\label{eqa}
\begin{aligned}
    \ln \mathcal{L} &= P(\log \rho_{{\rm out},i}|\{k_j\},b) \\
    &= \int d\rho_{{\rm in},i}P(\log \rho_{{\rm out},i} | \log \rho_{{\rm in},i}) P(\log \rho_{{\rm in},i}|\{k_j\},b).    
\end{aligned}
\end{equation} 

According to the joint probability map derived in Section~\ref{sec:map}, 
\( P(\log \rho_{{\rm out},i} | \log \rho_{{\rm in},i}) \) is well approximated by a Gaussian distribution {(denoted by $\mathcal{N}$)} across a wide range of \( \rho_{\rm in} \), even when \( \log \rho_\mathrm{in} \gtrsim 0 \) {(e.g., cases with $\log \rho_\mathrm{in} = 0$ and $\log \rho_\mathrm{in} = 2$ are presented in the lower-right panel of Figure~\ref{fig:mle})},
and thus it can be described by the form of
\begin{equation}\label{eqb}
    \log \rho_{{\rm out},i} | \log \rho_{{\rm in},i} \sim \mathcal{N}(\xi_i, [\Delta \xi_i]^2).
\end{equation}
The expectation $\xi_i$ and the uncertainty $\Delta \xi_i$ are both functions of $\log\rho_{{\rm in},i}$ and are defined as
\begin{equation}
\begin{gathered}
    \xi_i = \xi(\log \rho_{{\rm in},i}) \equiv E(\log \rho_{{\rm out},i}|\log \rho_{{\rm in},i}),\ {\rm and} \\
    [\Delta \xi_i]^2 \equiv D(\log \rho_{{\rm out},i}|\log \rho_{{\rm in},i}).
\end{gathered}
\end{equation}
{where \( E(\log \rho_{{\rm out},i}|\log \rho_{{\rm in},i}) \) denotes the conditional expectation and \( D(\log \rho_{{\rm out},i}|\log \rho_{{\rm in},i}) \) is the corresponding conditional variance.
}

In this regression, the residuals are normally distributed, and the probability of $\rho_{{\rm in},i}$ obeys
\begin{equation}\label{eqc}
\begin{gathered}
    \log \rho_{{\rm in},i}|\{k_j\},b \sim \mathcal{N} (\mu_i, \sum_j (k_j\Delta X_{j,i})^2 + \epsilon^2 ),
\end{gathered}
\end{equation}
where $\mu_i= \sum_j k_jX_{j,i} + b + \log(1+z_i) - \log(\text{baseline}_i)$. 
Here, we have included the intrinsic dispersion $\epsilon$ of the timescales, {where \(\epsilon \sim \mathcal{N}(0, \sigma_\epsilon^2)\).}

To proceed, we assume that the variance of $\log \rho_{{\rm in},i}$ is small so that $\xi_i$ can be replaced with a linear expansion around the mean value when convolving the probability density functions in Equation~\eqref{eqa}: 
$\xi_i=\xi(\mu_i)+\xi'(\mu_i)\cdot(\log \rho_{{\rm in},i}-\mu_i)$. 
Therefore, from Equations~\eqref{eqa},~\eqref{eqb} and~\eqref{eqc}, the likelihood function is given by
\begin{equation}\label{eq:like}
    \ln \mathcal{L} = -\frac12\left (\sum_i\frac{[\xi(\mu_i)-\log \rho_{{\rm out},i}]^2}{s_i^2} + \ln(2\pi s_i^2)\right),
\end{equation}
with $s_{i}^2 =  [\Delta \xi(\mu_i)]^2 + [\xi'(\mu_i)]^2 [\sum_j (k_j\Delta X_{j,i})^2 + \epsilon^2]$.
The variance $s_{i}^2$ represents the propagation of uncertainties from $X_{j,i}$ to $\rho_{\rm out}$.

From the probability map in Figure~\ref{fig:mle}, we infer that the derivative \( \xi'({\log \rho}) \) is equal to 1 on the left side and 0 on the right side. To evaluate the likelihood function, we employ a smoothed function as an approximation to characterize this feature:
\begin{equation}
        \xi'({\log \rho}) = 1 - \frac{1}{1 + {\rm exp}(-\frac{{\log \rho} + 0.51}{0.42})},
\end{equation}
where the parameters are fitted based on the simulation results in Section~\ref{sec:map}.

Following the user guide of \textsc{EMCEE} \citep{foreman-mackey_emcee_2013a} with the Equations~\eqref{eqo}~and~\eqref{eq:like}, the data can be well fitted to derive the dependence without calibrating the timescale for each observation. Our BADDAT method and its implementations are available on Github (\url{https://github.com/xiaris/BADDAT}).

\subsection{A Toy Model Illustrating Effectiveness of BADDAT}
\begin{figure}
    \centering
    \includegraphics[scale=0.42]{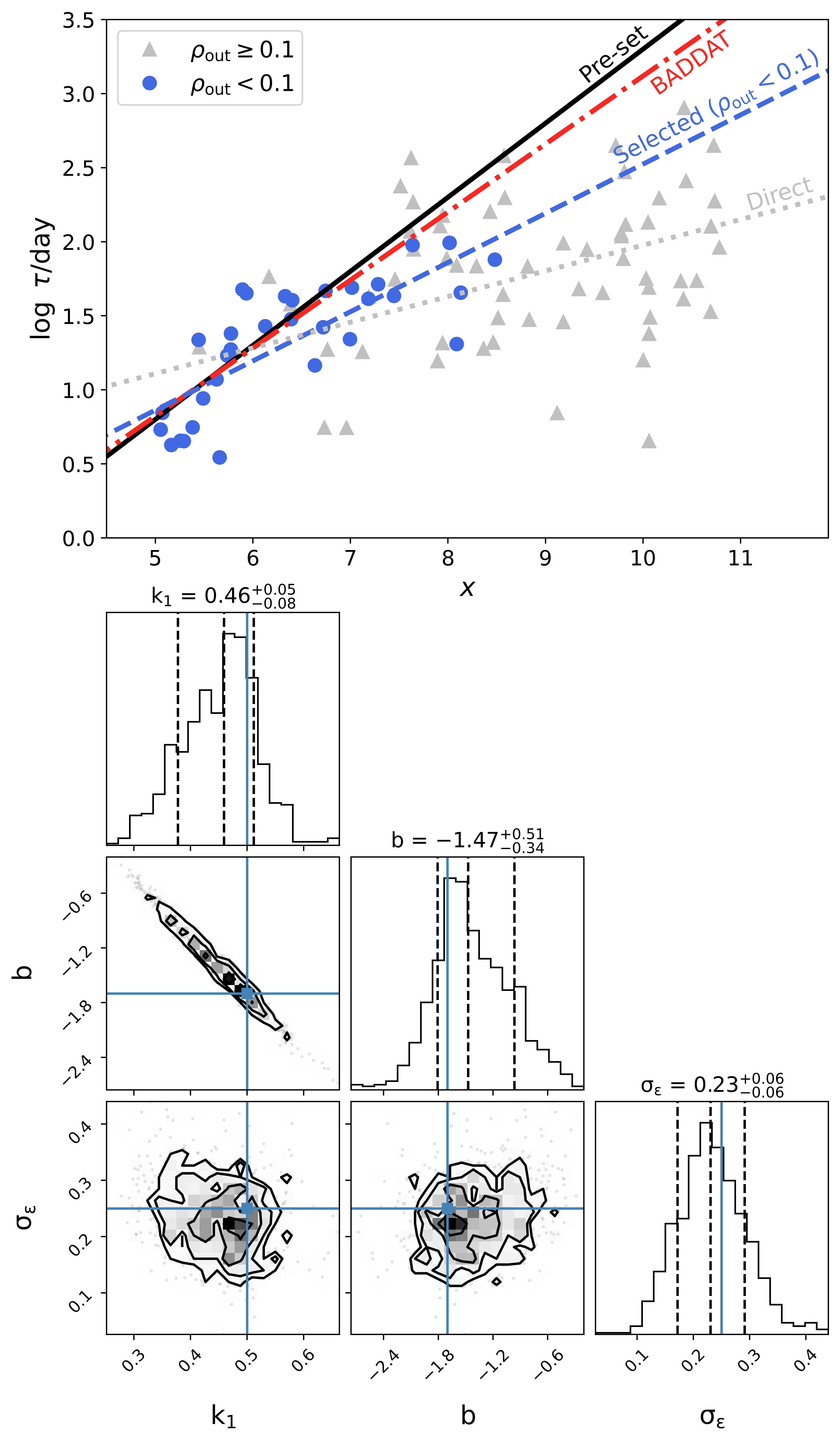}
    \caption{
        An illustration of recovering the dependence of the variability timescale on an independent variable $x$ with BADDAT. 
        The upper panel displays the measured timescale as a function of $x$, where data points with timescales shorter than $0.1 \times$baseline are shown as blue dots and the others as silver triangles. 
        The black solid line represents the pre-set correlation: {$ \tau_{\rm in} = 0.50x - 1.70$}.
        After incorporating scatter and underestimation effects, a direct fit (silver dotted line) yields {$\tau_{\rm out} = 0.17x + 0.24 $.}
        A fit applied only to timescales satisfying $\tau_{\rm out} < 0.1 \times$baseline (blue dashed line) gives {$ \tau_{\rm out} = 0.33x - 0.80 $}.
        Finally, our BADDAT approach ({red} dash-dotted line) recovers $ \tau_{\rm in} = 0.46x - 1.47$, which is much closer to the input relation than other approaches {for} this mock sample.
        The lower panel presents the posterior probability distributions of the fitted coefficient $k_1$, intercept $b$, and the scatter $\epsilon$ and their covariance, obtained using our {BADDAT} approach with \textsc{EMCEE}. 
        {The pre-set values and the 16th, 50th, and 84th percentiles of the posterior are indicated by blue solid and black dashed lines, respectively.}
    }
    \label{fig:rhc}
\end{figure}

To illustrate the effectiveness of BADDAT, we hypothesize an alternative correlation as a pilot study:
\begin{equation}\label{eq:0p5}
    \log (\frac{\tau_{\rm in}}{\rm days}) = 0.50 x - 1.70+\epsilon,
\end{equation}
where the independent variable $x$ follows a uniform distribution ranging from 5 to 11 {(representing a typical parameter such as the black hole mass)}, with a sample size of 100, and the intrinsic scatter $\epsilon$ follows a Gaussian distribution with a standard deviation of 0.25.
The values of $\tau_{\rm out}$ are generated based on the conditional probability function $P(\tau_{\rm out}|\tau_{\rm in})$, with the baseline randomly sampled from a uniform distribution between 10 and 1000. The number of sampled points within the baseline is also randomly drawn from a uniform distribution in the range of 10 to 1000.
As an indication, this sampling is intentionally kept simple and is comparable to or worse than that of the real observational samples considered in Sections~\ref{sec:test} \citep{2021Sci...373..789B,2024ApJ...975..160R}.  
Figure~\ref{fig:rhc} illustrates the effectiveness of {BADDAT} with the mock sample. 
For comparison, the data are fitted both directly and after applying the selection criterion $\tau_{\rm out} < 0.1 \times {\rm baseline}$.
The recovered correlation is much closer to the pre-set one compared to other approaches, indicating that our method significantly enhances the measurement of parameter dependence.
This improvement will be demonstrated in detail in Section~\ref{sec:test} using mock light curves based on specific samples.

\section{Recovering Hypothesis Correlations on Mock Light Curves}\label{sec:test}

\begin{figure}
    \centering
    \includegraphics[scale=0.33]{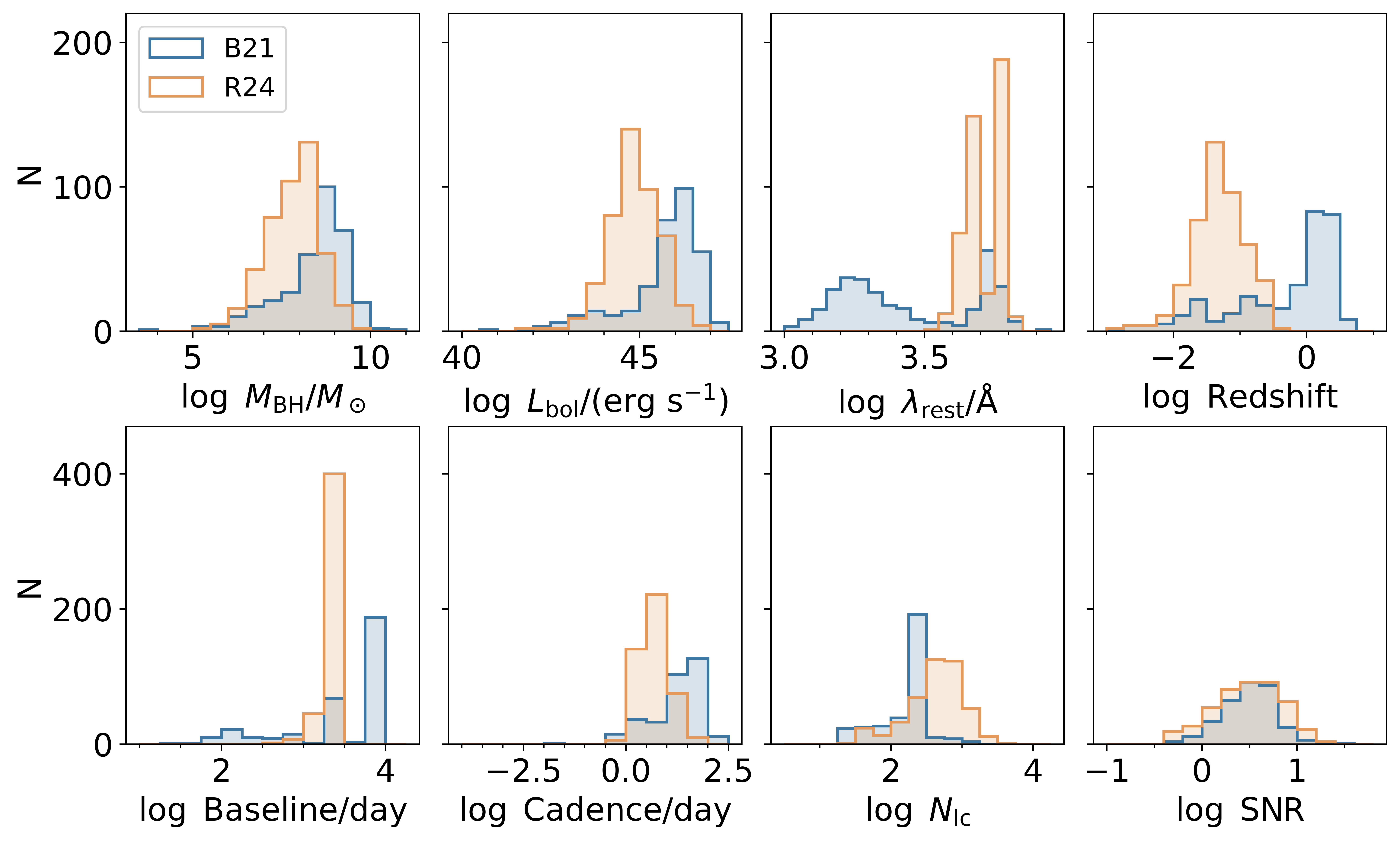}
    \caption{Distributions of black hole masses, luminosities, rest-frame wavelengths, {redshifts, baselines}, cadences, numbers of data points, and SNRs for the sources and light curves in the \burke\ and \ren\ samples.}
    \label{fig:sample}
\end{figure}

{
As a regression method, the effectiveness of {BADDAT} is highly dependent on the quality of the samples.
For convenience, we generate light curves based on real samples, including those from \citet{2021Sci...373..789B} and \citet{2024ApJ...975..160R} {(denoted as \burke\ and \ren\ hereafter)}, with the aim of demonstrating the recoverability of the dependence on specific samples.
We have excluded those with missing or unreliable measurements of black hole mass and luminosity. 
The bad fittings where $\tau_{\rm obs} < {\rm cadence}$ have been excluded due to the lack of knowledge in this region.
After the exclusions, \burke\ and \ren\ consist of 328 and 454 light curves corresponding to 308 and 239 sources, respectively.
An additional sample combining the two samples is constructed, with 782 light curves and 547 sources.
The distributions of black hole mass, luminosity, redshift, rest-frame wavelength, baseline, and cadence for the two samples are shown in Figure~\ref{fig:sample}.
}

The mock light curves are generated following the real sampling patterns of the observed light curves, i.e., preserving the same baseline, sampling cadence, and measurement errors while only replacing the observed magnitudes with DRW series characterized by a pre-set relation:  
\begin{equation}
\log (\frac{\tau_{\rm in}}{\rm days}) = 0.50 \log (\frac{M_{\rm BH}}{M_\odot}) - 1.70 + \log(1+z) + \epsilon, 
\end{equation}
where $\epsilon$ represents an assumed intrinsic scatter following a Gaussian distribution with a standard deviation of 0.25.

The simulations are repeated 500 times for each light curve, and the resulting mock light curves are fitted to the DRW model using the MLE method. 
Poor fits, where the fitting failed (\(\Delta \text{AIC}_{\rm low}<1\) or \(\Delta \text{AIC}_{\rm hi}<1\)) or the fitted timescale falls below the cadence, are excluded again.
For each mock light curve, we randomly select one surviving fit from the 500 simulations as the estimated timescale $\tau_{\rm out}$ to construct the mock sample.
We fit the data using \textsc{\hbox{EMCEE}} with the likelihood function ({i.e.,} Equation~\eqref{eq:like}) to derive the dependence of the parameters as the recovered result.

\begin{figure}
    \centering
    \includegraphics[scale=0.31]{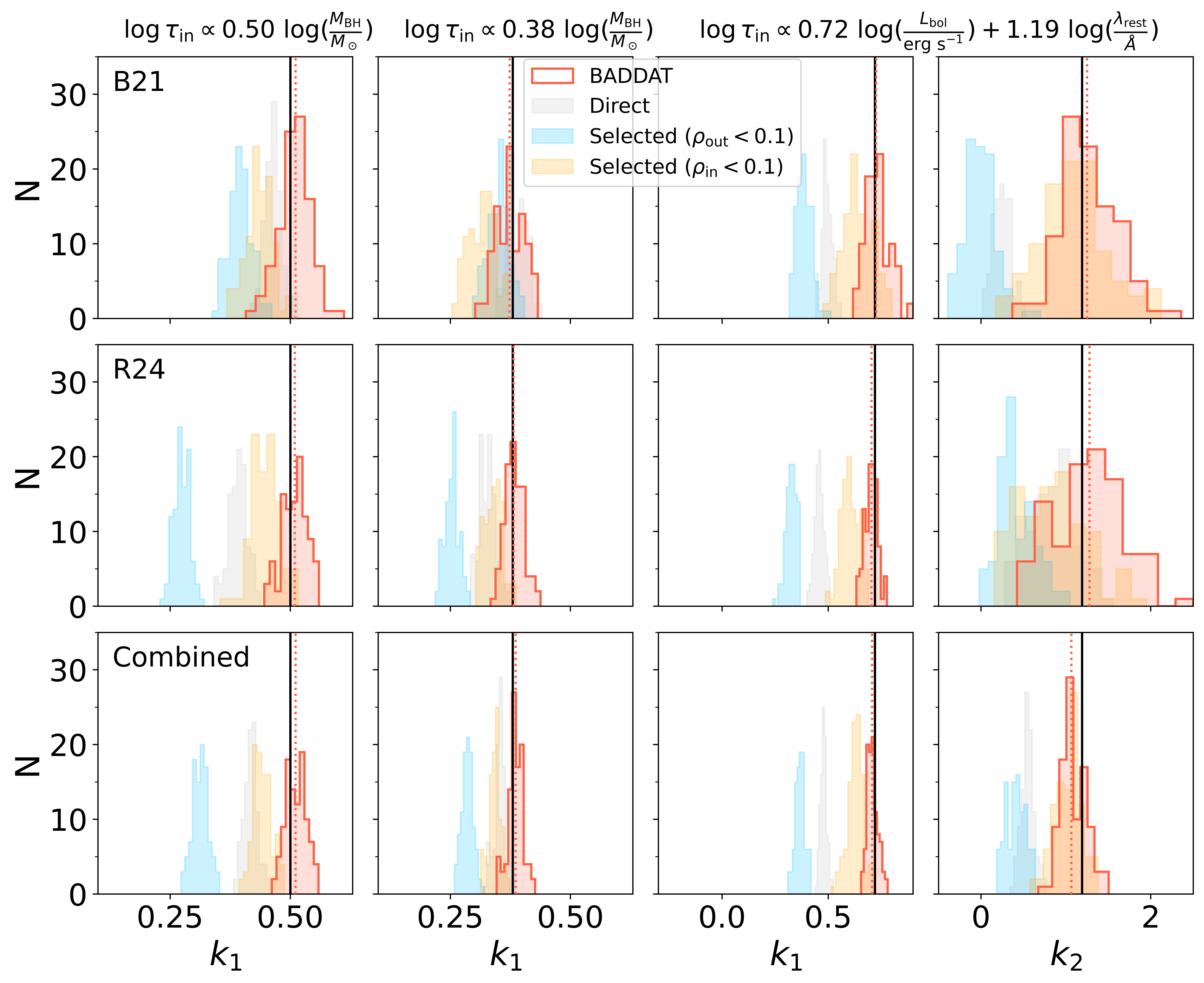}
    \caption{The distributions of the fitted coefficients for the timescale dependence on the parameters from different approaches {on the mock sample}.
    The vertical black solid line and red dotted line represent the pre-set coefficient and the median value of the distribution obtained from {BADDAT}, respectively.}
    \label{fig:simu}
\end{figure}

We then repeated this process 100 times to derive 100 mock samples, with $\tau_{\rm in}$ determined by several correlations as follows.
In addition to the alternative correlation presented {as} Equation~\eqref{eq:0p5}, we also consider several empirical correlations from the literature.
The correlation found by \citet{2021Sci...373..789B} can be expressed as:
\begin{equation}
    \log (\frac{\tau_{\rm in}}{\rm days}) = 0.38\log (\frac{M_{\rm BH}}{M_\odot}) - 1.01 + \log(1+z) + \epsilon,
\end{equation}
where $\epsilon$ is defined the same as in Equation~\eqref{eq:0p5}.
To test the effectiveness in two-variable cases, we consider the correlation found by \citet{2024ApJ...975..160R}:
\begin{equation}
    \begin{aligned}
        \log (\frac{\tau_{\rm in}}{\rm days}) &= 0.72\log (\frac{L_{\rm bol}}{{\rm erg\cdot s^{-1}}}) + 1.19\log(\frac{\lambda_{\rm rest}}{\text{\AA}}) \\
        &\quad - 34.2 + \log(1+z) + \epsilon.
    \end{aligned}
\end{equation}

\begin{table}
    \setlength{\tabcolsep}{5pt}
    \centering
    \begin{tabular}{ccccccccc}
    \hline
    \hline
        Equation~\eqref{eq0} & \vline & $k_1$ & &$b$ & ${\sigma_\epsilon}$\\
        \hline
\burke   &&$ 0.53 ^{+0.05}_{-0.04}$ &&$ -1.72 ^{+0.30}_{-0.36}$&$ 0.22 ^{+0.06}_{-0.06}$\\
\ren    &&$ 0.45 ^{+0.04}_{-0.04}$ &&$ -1.18 ^{+0.34}_{-0.34}$&$ 0.58 ^{+0.03}_{-0.03}$\\
Combined  &&$ 0.51 ^{+0.03}_{-0.03}$ &&$ -1.62 ^{+0.23}_{-0.25}$&$ 0.52 ^{+0.03}_{-0.02}$\\

         \hline
        Equation~\eqref{eq1} & \vline & $k_1$ & $k_2$ & $b$ & ${\sigma_\epsilon}$\\
        \hline
\burke   &&$ 0.58 ^{+0.07}_{-0.06}$ &$ 0.57 ^{+0.47}_{-0.39}$&$ -25.61 ^{+3.77}_{-4.90}$&$ 0.28 ^{+0.06}_{-0.06}$\\
\ren   &&$ 0.58 ^{+0.05}_{-0.05}$ &$ -0.33 ^{+0.52}_{-0.49}$&$ -22.60 ^{+3.09}_{-3.34}$&$ 0.51 ^{+0.03}_{-0.03}$\\
Combined   &&$ 0.59 ^{+0.03}_{-0.03}$ &$ 0.10 ^{+0.22}_{-0.24}$&$ -24.20 ^{+2.01}_{-2.08}$&$ 0.48 ^{+0.03}_{-0.03}$\\

    \hline
    \hline
    \end{tabular}
    \caption{Results from BADDAT regressions based on real light curves from samples {of} \burke, \ren, and their combination. {For comparison, \citet{2021Sci...373..789B} reported a coefficient of $k_1 = 0.38$ on $M_{\rm BH}$, whereas \citet{2024ApJ...975..160R} obtained $k_1 = 0.72$ on $L_{\rm bol}$ and $k_2 = 1.19$ on $\lambda_{\rm rest}$.}}
    \label{tab}
\end{table}

The red histograms in Figure~\ref{fig:simu} represent the {distributions} of recovered coefficients in specific samples using {BADDAT}, compared with the results from linear regression using two different selection approaches (blue and orange histograms) and those obtained without any selection (gray histograms).
As emphasized by \citet{2024ApJ...966....8Z} and \citet{2024ApJ...975..160R}, we further demonstrate that the selection {of} $\tau_{\rm in} < 0.1\times\text{baseline}$ performs much better than the other selection, $\tau_{\rm out} < 0.1\times\text{baseline}$, which is even worse than direct fitting without selection.
However, this selection approach ($\tau_{\rm in} < 0.1\times\text{baseline}$) strongly assumes that the exact dependence is known in advance and requires a high-quality, sufficiently large sample.
In comparison, by relaxing the requirements and incorporating additional information {contained in underestimated timescales}, our new BADDAT approach achieves better compatibility and is nearly unbiased, outperforming both selection approaches across the {examined} samples.

Furthermore, compared to individual samples, the results from the combined mock sample are more precise (with lower uncertainties), indicating that, although the quality of the observations does not significantly improve, enlarging the sample enhances the estimation of the dependence of variation properties on physical parameters.

\section{Discussions}\label{sec:imp}

Based on the knowledge derived above, the dependence of the timescale {on AGN physical properties} can be well estimated from the samples.
{In this work}, we apply {BADDAT to the light curves of \burke, \ren, and their combination, in order to revisit the previously reported correlations} \citep{2021Sci...373..789B,2024ApJ...975..160R}.

Considering the empirical relations described in Section~\ref{sec:test}, we fit the equations:
\begin{equation} \label{eq0}
    \log (\frac{\tau_{\rm in}}{\rm days}) = k_1 \log (\frac{M_{\rm BH}}{M_\odot}) + b + \log(1+z) + \epsilon,
\end{equation}
and
\begin{equation} \label{eq1}
    \log (\frac{\tau_{\rm in}}{\rm days}) = k_1 \log (\frac{L_{\rm bol}}{\rm erg\cdot s^{-1}}) + k_2 \log(\frac{\lambda_{\rm rest}}{\text{\AA}})+ b + \log(1+z) + \epsilon.
\end{equation}
The fitting results are shown in Table~\ref{tab}.  
We found that, in both correlations, the dependence of timescale on black hole mass or bolometric luminosity is consistent across different samples, whereas the dependence on rest-frame wavelength shows greater deviation with larger uncertainties.
Note that the light curves with {low SNRs ($\text{SNR} <1$)} are included. 
However, we confirm that the results show negligible changes when applying an additional selection criterion requiring $\text{SNR} > 1$, as also indicated by our simulations.

The best fit for the dependence on $M_{\rm BH}$ is close to 0.5 within about $1\sigma$ uncertainty, which is consistent with a thermal or orbital timescale as predicted by the static standard accretion-disk theory \citep{1973A&A....24..337S}.
For the dependence on $L_{\rm bol}$ and $\lambda_{\rm rest}$, \citet{2024ApJ...966....8Z} employed the corona-heated accretion-disk reprocessing model \citep{2020ApJ...902....7S} to derive a relation expressed as $\tau \propto M_{\rm BH}^{0.65} \dot{m}^{0.65} \lambda_{\rm rest}^{1.19} \propto L_{\rm bol}^{0.65} \lambda_{\rm rest}^{1.19}$.
{Our results support a similar dependence on $L_{\rm bol}$, but the wavelength dependence is weaker than predicted and shows inconsistencies across different samples.
According to our simulations, if a strong dependence exists between the timescale and the rest-frame wavelength, it should still be discernible from the combined sample despite its large uncertainties.
Therefore, we infer that the dependence of the variation on the wavelength may not follow a universal form and could vary across different samples.}

This work is based on the assumption that the DRW model provides a reliable description of intrinsic light curves. 
However, previous studies have reported deviations from the DRW process.
For example, deviations at shorter timescales have been observed in both supermassive black holes (e.g., the {\it Kepler} AGNs; \citealt{Mushotzky2011ApJ}) and intermediate-mass black holes (e.g., NGC 4395; \citealt{2024ApJ...969...78S}), highlighting the need for a careful assessment when applying the DRW model, as its validity in these cases may be uncertain.  
Furthermore, the effectiveness of {BADDAT} is directly determined by the joint probability distribution map, which still has potential for improvement.
Nonetheless, the strong coherence between the DRW timescale and both the black hole mass and luminosity in the samples used in this work, along with the consistency across the samples, ensures the fundamental effectiveness of {BADDAT} in these specific cases.
Therefore, as recommended, performing simulations on each sample prior to application would be beneficial.

The {ongoing WFST \citep{2023SCPMA..6609512W} and LSST \citep{2019ApJ...873..111I}} surveys are expected to yield extensive AGN variability data. 
As {BADDAT} relaxes the selection criteria and offers a nearly unbiased framework for future studies, the sample size will be greatly increased, and the estimation of the dependencies of the variability timescale on physical parameters will be further improved.

\section*{Acknowledgements}

{We sincerely thank the anonymous referee for the careful review and helpful comments, which have significantly improved this work.}
This work is supported by the National Key R\&D Program of China (grant No. 2023YFA1608100) and the NSFC grants (12025303).
R.S.X gratefully acknowledges the support of Cyrus Chun Ying Tang Foundations.

\section*{Data Availability}

Our BADDAT package is publicly available at \url{https://github.com/xiaris/BADDAT}. 
The data used in this work are sourced from \citet{2021Sci...373..789B} and \citet{2024ApJ...975..160R}. 
The analysis in this work utilizes the following software: 
\textsc{celerite} \citep{celerite}, 
\textsc{corner} \citep{corner}, 
\textsc{emcee} \citep{foreman-mackey_emcee_2013a}, 
\textsc{matplotlib} \citep{Hunter:2007}, 
\textsc{numpy} \citep{harris2020array}, 
\textsc{scikit-learn} \citep{scikit-learn}, 
\textsc{scipy} \citep{2020SciPy-NMeth}, and
\textsc{taufit} \citep{2021Sci...373..789B}.



\bibliographystyle{mnras}
\bibliography{BADDAT} 








\bsp	
\label{lastpage}
\end{document}